\documentclass{bvm}

\makeatletter
\let\blx@rerun@biber\relax
\makeatother

\begin{document}

\newcommand{\bvmyear}{2023}

\selectlanguage{english}

\title{Towards clinical translation of deep-learning based classification of DSA image sequences for stroke treatment}

\titlerunning{DSA classification for stroke treatment}

\author{
	Timo \lname{Baumgärtner} \inst{1}, 
	Benjamin~J. \lname{Mittmann} \inst{1}, 
	Till \lname{Malzacher} \inst{2},
	Johannes \lname{Roßkopf} \inst{2},
	Michael \lname{Braun} \inst{2}, 
	Bernd \lname{Schmitz} \inst{2}, 
	Alfred~M. \lname{Franz} \inst{1}
}

\authorrunning{Baumgärtner et al.}

\institute{
\inst{1} Department of Computer Science, Ulm University of Applied Sciences\\
\inst{2} Neuroradiology Section, District Hospital Guenzburg 
}

\email{alfred.franz@thu.de}

\maketitle

\begin{abstract}
In the event of stroke, a catheter-guided procedure (thrombectomy) is used to remove blood clots. Feasibility of machine learning based automatic classifications for thrombus detection on digital substraction angiography (DSA) sequences has been demonstrated. It was however not used live in the clinic, yet. We present an open-source tool for automatic thrombus classification and test it on three selected clinical cases regarding functionality and classification runtime. With our trained model all large vessel occlusions in the M1 segment were correctly classified. One small remaining M3 thrombus was not detected. Runtime was in the range from 1 to 10 seconds depending on the used hardware. We conclude that our open-source software tool enables clinical staff to classify DSA sequences in (close to) realtime and can be used for further studies in clinics.
\end{abstract}

\section{Introduction}
Worldwide,  ischemic stroke, in which a blood clot blocks blood vessels in the brain, is one of the most common causes of death~\cite{3320-lopez}. In addition to drug treatment (thrombolysis), removal of the blood clot using a catheter-guided procedure (thrombectomy) has now become widely accepted and has shown a significantly better outcome for patients~\cite{3320-goyal}. However, there continue to be numerous challenges in performing thrombectomy, resulting in reperfusion of occluded vessels being achieved in only 70-80\% of cases, with treatment remaining unsuccessful in the remaining patients~\cite{3320-yoo}.

Thrombectomy is usually performed under fluoroscopic guidance. Here, digital subtraction angiography~(DSA) can be used to visualize the vascular tree in relation to the instruments.
Vascular occlusions frequently occur in the middle cerebral artery~(MCA), specifically in the M1 - M3 segments, with the MCA having a diameter of approximately 3~mm in the M1 segment immediately after the branch from the internal carotid artery, which then narrows further in the M2 segment and M3 segment. M3 occlusions are more difficult to detect and treat, but are also often less critical than, for example, M1 occlusions.
One challenge for the physician is to quickly identify whether a blood clot is still present in the DSA sequence just acquired or whether it has already been successfully removed.
Automatic classification of DSA sequences using machine learning methods has already been demonstrated in studies, for example by Nielsen~et~al. who used an EfficientNet-B0-based Convolutional Neural Network (CNN) with Gated Recurrent Units~(GRU) to classify with respect to Thrombolysis In Cerebral Infarction~(TICI) and achieved an accuracy of $ 0.95 \pm 0.03 $~\cite{3320-nielsen}. Another work by Su~et~al. uses a multi-path CNN for automatic TICI classification with which they achieve an average area under the curve~(AUC) value of $0.81$~\cite{3320-su}. In a preliminary work of ours, we demonstrated that using an EfficientNetV2 architecture with GRU, DSA sequences can be classified into thrombus-free and non-thrombus-free, achieving an AUC of $0.94$~\cite{3320-mittmann}. 

While the feasibility of automatic classification has been demonstrated by the listed studies, to the best of our knowledge, the classifications were performed offline using dedicated computers and could not yet be used live in the clinic.
With the work presented here, based on our preliminary work, we want to go one step further towards clinical translation: (1)~we present an open-source application that can be installed together with the trained network on arbitrary computers and classify DICOM data there, (2)~after performing a 5-fold cross-validation in the preliminary evaluation work~\cite{3320-mittmann}, we re-train the network with all training data in this work, and (3)~test the classification on new datasets acquired in the clinic with a different DSA system. The trained network and the classification tool are published.

\section{Methods}
\label{3320-section-methods}
\subsection{Open-source application for classification}
Focusing on enabling users from medical research in using the classification, we develop a prototypical graphical user interface. We envision to provide several functionalities, such as (1)~Loading of DICOM and nifti image files, (2)~Showing an one-image-preview of the loaded series, (3)~Loading of any model that shall be used for classification, (4)~Selection of the classification threshold based on statistical evaluation on the training data and displaying these statistics, (5)~Classification of the loaded data and (6)~Keeping the requirements of the application in terms of the hardware as low as possible with a classification result provided in less than 60 seconds.

In order to combine the graphical user interface (GUI) capabilities of the .NET framework using Windows Presentation Foundation~(WPF) and the typical python-based PyTorch implementation for machine learning, we employ a classic client-server architecture. The python service can therefore be started as the server, providing an interface that the user application can query. These endpoints can provide functionality like model interference (e.g. classifying provided images) or image rendering, to employ the same image preprocessing as it is used for the model. We use the python package nibabel (\url{https://nipy.org/nibabel/}) to load the nifti-files. In case of DICOM data, we use the commandline tool plastimatch (\url{https://plastimatch.org/}) to convert them into the nifti file format. In order to comply with the security of patient data and to keep loading times short, both the python service as well as the user application need to run on the same machine, sharing the same filesystem.

In order to make the model applicable for non-technical users such as clinical staff, we provide an interactive view, enabling to choose the threshold (e.g. the value at which the application labels a sample as positive/negative) and run the classification. Thereby, one can optimize different metrics (like the Matthews Correlation Coefficient MCC~\cite{3320-chicco}) according to their needs. We initially provide an optimized threshold value of 0.57, which is found by optimizing the rates of false positives and true positives. We therefore adapt the proposed closest-point-criterion~\cite{3320-perkins}, which, when using $ FPR $ instead of $ (1-FPR) $ as abscissa, can be rewritten as
\begin{align} 
	 \text{threshold} = \arg \max_{t} \sqrt{(1-{FPR}_{t})^2+{TPR}_{t}^2}
\end{align}
where $ FPR $ is the false positive rate and $ TPR $ is the true positive rate, found by evaluating the correctness of classification results utilizing threshold $ t $ on the training data.

\subsection{Network architecture and model retraining}
One network setup that yielded good results in~\cite{3320-mittmann} is EfficentNetV2+GRU. We employ this architecture for classification, as shown in Fig.~\ref{3320-fig-network-schema}. We utilize two models, one that was trained for frontal sequences, and another that was trained for lateral sequences. The final classification is the mean of both predictions.

In~\cite{3320-mittmann}, crossfold-validation was used to estimate the quality of the classification, since the amount of data samples was quite limited. In order to provide maximum performance of the model, we combined training and test dataset and trained one single model on all of it. We use the metrics of the crossfold-validation to estimate the models performance. However, we plan to continually assess the performance of our model with future data, as discussed in detail in Sec.~\ref{3320-section-discussion}. 

\subsection{Evaluation methods}
The objective was to evaluate the software and the new trained model regarding (1)~functionality, (2)~classification correctness and (3)~runtime on different systems. We were able to evaluate our system on three new cases that were generated with another angiography unit than the training data, namely a \emph{Siemens ARTIS icono biplane} instead of a \emph{Siemens ARTIS zee biplane}. These three cases were selected by physicians, consisting of two DSA sequences each: One before the treatment was started, inherently containing one or even multiple thrombi. The other one was taken after the treatment. The first case shows an M1 thrombus, all thrombi were removed. The second case initially shows an M1 thrombus which was removed, one small peripheral M3 thrombus was remaining. The third case shows a thrombus in the carotid terminus, all thrombi were successfully removed, however, the sequence shows flow reversal in the vessels, which might be mistaken for an occlusion by inexperienced observers.

We benchmarked several different hardware configurations, solely running on CPU or utilizing GPU, and both on various budget levels. For testing the CPU performance, we compare an Intel i7-7700K with 8 threads with an Intel i7-11700KF with 16 threads. Regarding the GPU, we evaluate a Nvidia GTX 1050Ti Mobile and a Nvidia RTX 3090. 
The tests were conducted under Python 3.10.8, PyTorch 1.18, CUDA 11.7.1 and Nvidia GPU driver 526.98, which were the most recent compatible versions at this time.
We therefore benchmarked the three new cases, resulting in $ n=6 $ datapoints, and took the mean of the timings. Since the classification time might deviate depending on the file size, we also calculated the standard deviation for these measurements.

\section{Results}
\label{3320-section-results}
\begin{figure}[b]
	\begin{subfigure}{0.35\textwidth}
		\includegraphics[width=\linewidth]{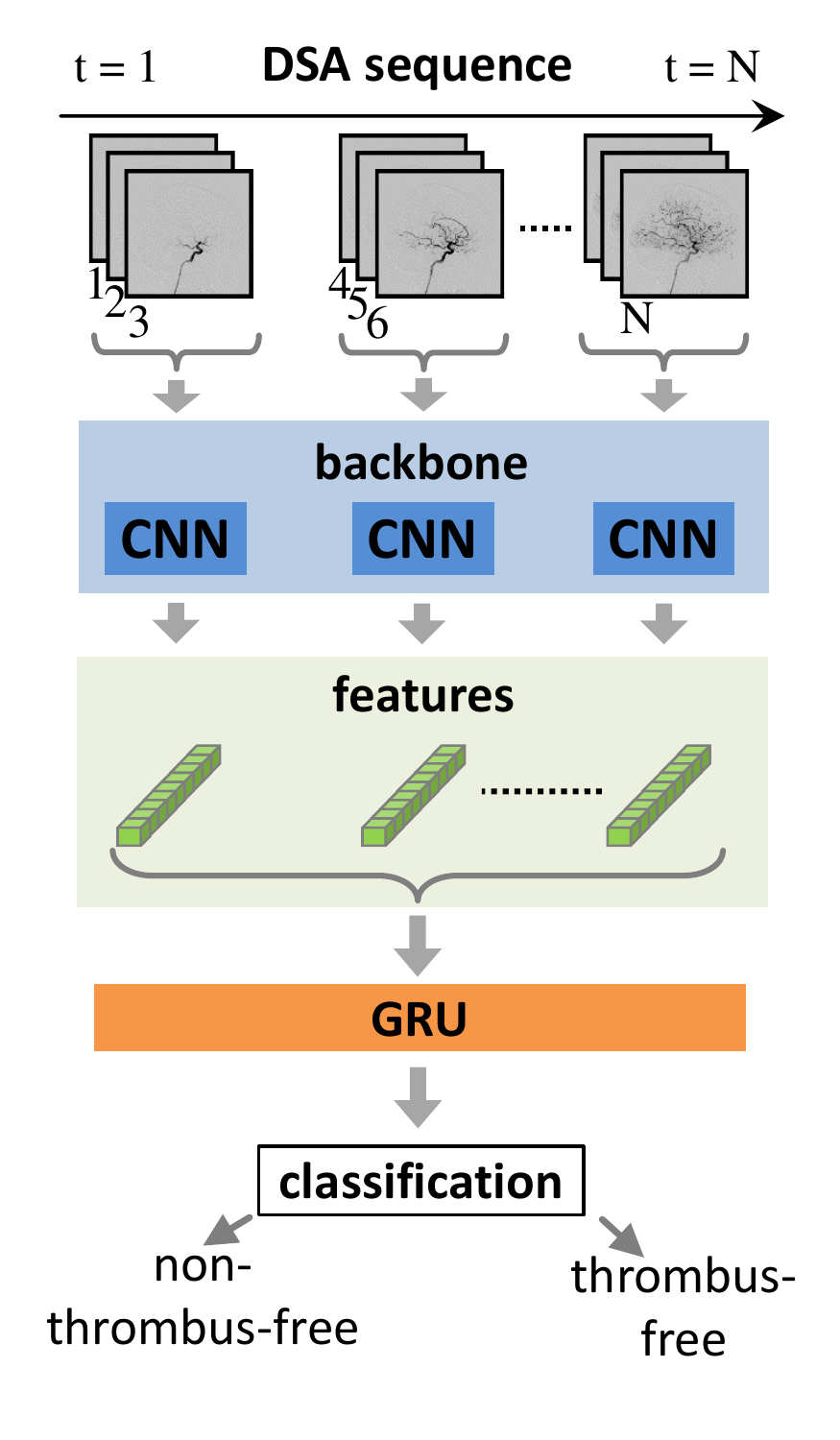}
		\caption{Architecture of the neural network (adapted from~\cite{3320-mittmann}).}
		\label{3320-fig-network-schema}
	\end{subfigure}
	\hfill
	\begin{subfigure}{0.62\textwidth}
		\includegraphics[width=\linewidth]{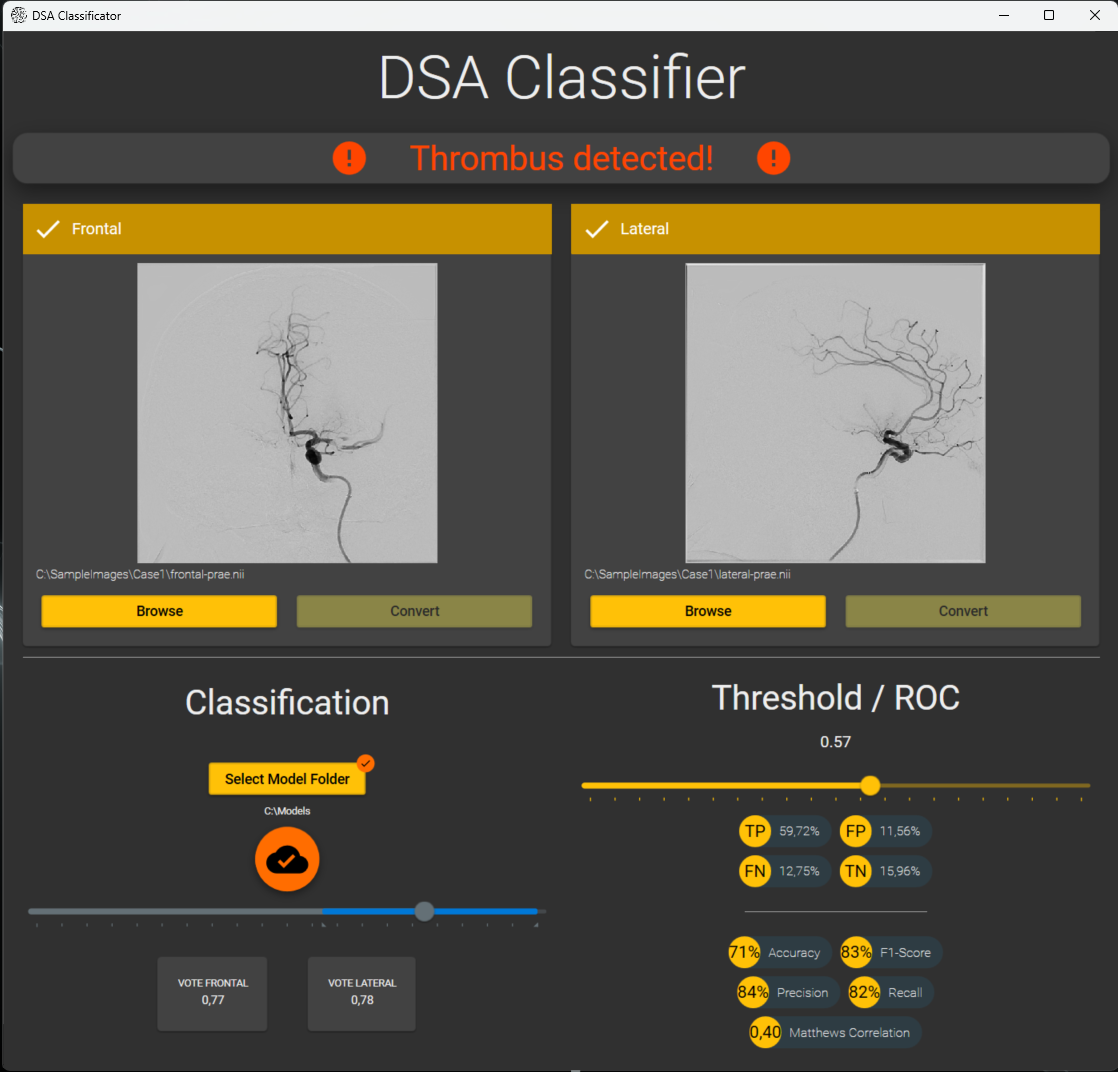}
		\caption{Screenshot of the application.}
		\label{3320-fig-screenshot-application}
	\end{subfigure}
	\caption{Details of the application.}
\end{figure}

A screenshot of our developed application is shown in Fig.~\ref{3320-fig-screenshot-application}.
In order to support the concept of Open Science, we provide our application as well as the trained model to the community by making it open-source (\url{https://osf.io/n8k4r/}).

In all three cases, the first and second sequences were correctly classified as non-thrombus-free and thrombus-free. However, the small remaining M3 thrombus was not detected, a behavior that we already know from our previous study which needs discussion (see Sec.~\ref{3320-section-discussion}). In case 3, the raw output ($0.49$) was close to the threshold of $0.57$, but still below, which was due to a misleading flow reversal in the vessels, by which experienced physicians would not have been irritated. Tab.~\ref{3320-tab-cases} shows the raw combined classification result for all three cases along with the respective file size of the frontal and lateral sequence combined. 

\begin{table}[t] 
	\caption{Details about the tested cases.}
	\label{3320-tab-cases}
	\begin{tabular*}{\textwidth}{l@{\extracolsep\fill}llllll}
		\hline
		
		& \multicolumn{2}{c}{Case 1} & \multicolumn{2}{c}{Case 2} & \multicolumn{2}{c}{Case 3} \\
		& Pre & Post & Pre & Post & Pre & Post \\
		\hline
		Size in MB (Combined) & 283 & 199 & 199 & 210 & 210 & 263 \\
		Raw Classification Output & 0.77 & 0.24 & 0.70 & 0.25 & 0.79 & 0.49 \\
		
		\hline
	\end{tabular*}
\end{table}

The resulting benchmarks can be seen in Tab.~\ref{3320-tab-hardware}. The timings were calculated using six datapoints, namely our three cases that each consisted of preinterventional and postinterventional sequences. We show the approximate time in seconds needed to classify one case, in order to estimate necessary hardware constraints. Preliminary steps are not included, but typically add less than 15 seconds for data conversion (if necessary) and 5 seconds for image loading along with preprocessing.

\begin{table}[t]
	\caption{Required mean classification time in seconds with standard deviation (n=6).}
	\label{3320-tab-hardware}
	\begin{tabular*}{\textwidth}{l@{\extracolsep\fill}lll}
		\hline
		Highend GPU & Basic GPU & Highend CPU & Basic CPU \\
		Nvidia RTX 3090 & Nvidia GTX 1050Ti & Intel i7-11700KF & Intel i7-7700K \\
		24GB VRAM & 4GB VRAM & 32GB RAM & 16GB RAM  \\
		\hline
		1.1s $\pm$ 0.1 & 1.8s $\pm$ 0.1 & 4.8s $\pm$ 0.5 & 8.8s $\pm$ 0.9 \\
		\hline		
	\end{tabular*}
\end{table}

\section{Discussion}
\label{3320-section-discussion}
Despite the new angiography system, our system was able to classify all sequences correctly for large vessel occlusions in the M1 segment. In the second case, our system was unable to detect the remaining small peripheral thrombus. As these kind of thrombi were not focus of the training process, we already expected this behavior. However though, as these cases would usually not be treated by thrombectomy, this was not part of the goals for the system (also see~\cite{3320-mittmann}). In the future, we plan to extend the training data and retrain the model to investigate if M2 and M3 thrombi can also be detected.

We conclude, that for our tool to be used, no highend hardware is needed. If necessary, the classification can be done on the CPU in still reasonable time, although we see that the maximum number of threads influences performance.
However though, the system already greatly benefits from entry-level graphic cards. The use of high-end hardware in return does not yield great advances, when seen in relation to the costs. The classification time of up to approximately 10 seconds seems acceptable. In addition, the conversion of the DICOM data along with preprocessing is estimated in the experimental runs to take at most 30 seconds. Thus, the total duration also seems to be in an acceptable range, especially since the  runtime can certainly still be optimized.

As the values of the various metrics will likely be different on unseen new data, the use of these values for estimating an optimal threshold should be treated carefully. We therefore propose to incorporate ways of obtaining user feedback concerning the correctness of the classification, to store this feedback along with the raw classification output, and to continually recompute and optimize the threshold using these values. Furthermore, more test cases are needed. We currently prepare a larger study that assesses the quality of our model.

We envision a system that provides physicians hints right in the moment of the thrombectomy with an additional safety. For this, we plan a watch-dog-system that runs in the background and only warns the physician if it detects remaining thrombi. Since the current system can only classify whole sequences into thrombus-free and non-thrombus-free, we are working on visualizing the parts of the sequence that lead to the classification. This will give clinicians more and deeper insight into the decision, which in turn will make the system more credible and the process more efficient.

It’s important to note that this does not, at any time, replace the educated assessment of the physicians themselves, nor the ones of assisting staff. The goal is to provide an additional layer of assessment that solely provides a supplementary opinion that, in case of a warning, might motivate the physician to check for critical sections once more. By, in the best case, triggering a reassessment from the clinical staff, treatment quality will be at least the same as without the application.

We faced problems with the specific PACS system in the clinic, an \emph{AGFA HealthCare IMPAX EE R20 XVIII SU1}, as it sometimes failed to export DICOM data correctly. We therefore currently investigate several possibilities for a realtime connection, such as the connection to the angiography system's manufacturer's API as well as the potential use of framegrabbers. In addition to converting DICOM data on import, the current prototype also allows nifti files to be imported directly, which are useful for demonstration, research and fallback purposes.

Whereas the path to a medical product is still long, we are confident that our application along with the planned study is a step in the right direction. Also, we may help radiologists test the classification in their clinical environment, even during ongoing procedures if ethically approved.

\begin{acknowledgement}
This work was funded by the Federal Ministry for Economic Affairs and Climate Action (BMWK, Funding Code: ZF4640301GR8)
\end{acknowledgement}

\printbibliography

\end{document}